# Subsurface charge accumulation imaging of a quantum Hall liquid

S.H. Tessmer*, P.I. Glicofridis, R.C. Ashoori, and L.S. Levitov
*Department of Physics, Massachusetts Institute of Technology, Cambridge, Massachusetts 02139*

M.R. Melloch
*Department of Electrical Engineering, Purdue University, West Lafayette, Indiana 47907*

**The unusual properties of two-dimensional electron systems that give rise to the quantum Hall effect have prompted the development of new microscopic models for electrical conduction.[1-6] The bulk properties of the quantum Hall effect have also been studied experimentally using a variety of probes including transport[7,8], photoluminescence[9,10], magnetization[11], and capacitance[12,13] measurements. However, the fact that two-dimensional electron systems typically exist some distance (about 1000 Å) beneath the surface of the host semiconductor has presented an important obstacle to more direct measurements of microscopic electronic structure in the quantum Hall regime. Here we introduce a cryogenic scanning-probe technique-- 'subsurface charge accumulation' imaging-- that permits very high resolution examination of systems of mobile electrons inside materials. We use it to image directly the nanometer-scale electronic structures that exist in the quantum Hall regime.**

The subsurface charge accumulation (SCA) probe measures the local accumulation of charge in a two-dimensional electron system (2DES) in response to an applied ac excitation. We use it to examine the quantum Hall system over areas as large as several microns, and with a spatial resolution of ~900 Å. For magnetic fields near Hall plateaus, fine structures appear in the SCA images. These stuctures move and contort rapidly as the field is varied, even though they reappear on successive Hall plateaus.

In contrast to other techniques used to study a 2DES located beneath an insulating surface,[14, 15] SCA imaging is an ac method,[16-19] and has significant advantages for the study of the QHE over measurements which image the distribution of static charges. As samples contain static charges from both the surface and dopant layers underneath the surface, these regions can mask charging features in the 2D layer.[15] SCA imaging, operating at frequencies of around 100 kHz (very low compared to other ac methods), is uniquely suited for studying the accumulation of mobile charges in the 2DES in magnetic fields.

SCA microscopy of the 2DES proceeds as follows. A sharp conducting tip is positioned about 50 Å above the sample surface and connected to a highly sensitive charge detector (see Fig. 1). An ac excitation voltage is applied to the 2DES through an ohmic contact (2 mm away from the tip), and no other contacts are made to the sample. The excitation causes charge to flow in and out of the 2DES, in turn inducing charge to flow in and out of the tip. In scanning the tip and using synchronous lock-in detection of the induced charge, we obtain two images, one for charge accumulating in-phase ($Q_{in}$) and the other 90° (lagging) out-of-phase ($Q_{out}$) with the excitation. The microscope also operates in other modes allowing determination of surface topography (tunneling mode) and local accumulations of static charge (Kelvin probe mode).[20]

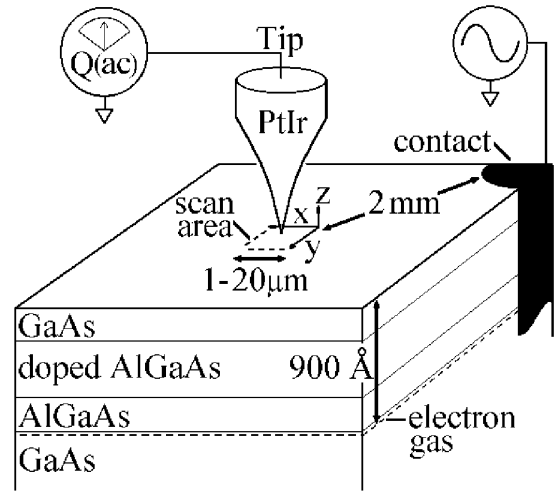

**Figure 1** Schematic of the sample and measurement configuration. The 2DES exists at the interface of $Al_{.3}Ga_{.7}As$ and GaAs 900 Å beneath the surface. The ac excitation applied to the 2DES causes this layer to charge and discharge due to the self-capacitance of the layer. A sharp metal tip is connected to a high electron mobility transistor (HEMT) with extremely high sensitivity to electrical charge[31] (0.01 *electrons/Hz$^{1/2}$*) and scanned 50 Å above the sample surface. SCA microscopy produces an image of charge accumulating in the 2DES by detecting charge induced on the tip due to charging in the layer. The spatial resolution is given approximately by the depth of the 2DES below the sample surface. The microscope operates with the sample immersed in liquid helium-3 (T=300 mK) and has a 20 mm long scan range. The static electron density in the 2DES is $n_b$ = 3.0 x $10^{11}$ $cm^{-2}$ and the transport mobility is m=450,000 $cm^2$/Vs. The small amplitude of the 100 kHz excitation applied to the 2DEG (1-4 mV) and the smallness of the capacitance between the tip and the 2DES (compared to the self-capacitance of the 2DES) ensure that the tip is practically non-perturbative of the 2D system.[26]

As a test of our microscope for studying quantum Hall features, we first use it to locally perturb the 2DES. If large voltages (>2 V) are applied to the tip, static and immobile charge can be deposited on the exposed surface of the sample or in the donor layer between the surface and the 2DES, thereby locally altering the electron density in the 2DES. Fig. 2 shows a series of six $Q_{in}$ (in-phase charging) SCA images centered on such a location. In these images, dark and bright represent small and large $Q_{in}$ signals respectively. Notice the dark spot which appears as the field is raised to 1.5 T and grows for larger fields.

The origin of the dark spot can be understood by considering basic properties of the 2DES in the quantum Hall state. Any region containing a density which precisely fills an integer number of Landau levels is "incompressible" and does not accumulate charge or conduct electricity. Such regions therefore appear dark in our images. At the center of the images in Fig. 2, the electron density has a broad minimum (with the density reduced to about ¼ of the bulk value). At the borders of the dark spot, the density is such that one Landau level is filled, preventing charge transport to the interior. Outside of the dark spot, the density of electrons is high enough so that more than one level is still filled, and charge accumulates freely.

Incompressible regions such as this dark spot appear in some QHE theories[5, 21, 22]. They are introduced to explain the transitions between the QH plateaus. At magnetic fields just below the quantum Hall plateaus the theory predicts that the 2DES contains a system of incompressible droplets. As in the case of Fig. 2, as the field is increased, these droplets grow and begin to merge. Thus, entering the Hall plateau, the 2DES undergoes a percolation transition with the network of merged droplets impeding conduction across the sample, and thus producing a quantum Hall plateau.[1, 4] However, in the absence of long range potential fluctuations, theories do not require the existence of droplets at fixed positions.[23-25] The local charging observed in Fig. 2 displays sharp contrast, consistent with the concept of distinct, insulating and conducting regions.

Having shown that the microscope can resolve QH features, we now use it to examine an unperturbed system. In contrast to the perturbed region, the SCA images only develop features over narrow ranges in magnetic field around integer Landau level filling (ν) values. As the magnetic field is adjusted toward the center of the Hall plateau, the charge accumulation appears to diminish roughly uniformly at all locations and detailed spatial patterns of charging appear.

Here, we concentrate on the ν=4 Hall plateau which is centered around 3.04 T in this sample. Images taken at fields of more than 0.15 T away from the center of the Hall plateau show no features. This indicates that spatial fluctuations in the charging (per volt of excitation amplitude) are less than $1 \times 10^{-18}$ coul./V. For comparison, the dark region observed at 3.20 T in the perturbed area (Fig. 2) displayed a reduction of $50 \times 10^{-18}$ coul./V. In contrast, clear and intricate charging features of amplitudes in the range of $20 \times 10^{-18}$ coul./V are observed for fields near the plateau center. The features are extraordinarily sensitive to magnetic field, with little correlation between the images at fields differing by 0.020 T (or 0.7%).

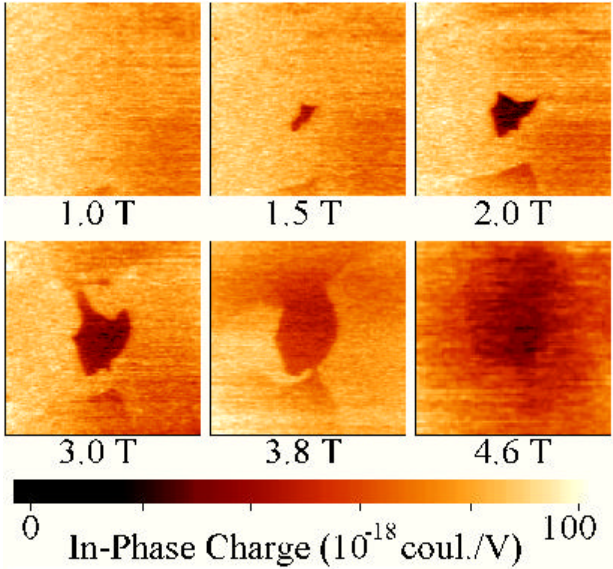

**Figure 2** In-phase ($Q_{in}$) SCA images (2.7mm x 2.7mm) of a region of perturbed electron density in the 2DES for various magnetic fields applied perpendicular to the 2D plane. Static charge measurements made in Kelvin probe mode show the 2D electron density has a shallow minimum at the center of the images and increases to the bulk value away from the center. For a magnetic field as high as 1.0 T, the SCA image exhibits few features. However at 1.5 T, the signal drops to a lower value when the tip is located near the center, displaying a small feature (dark spot) which grows with increasing magnetic field. The behavior can be explained if the spot is surrounded by a nonconducting boundary at an integer Landau level filling factor. The boundary of the spot marks the position in space for integer filling, and this position moves farther from the center to higher electron density regions as the field is raised. The density reduction at the spot center can be discerned by considering the magnetic field at which the black spot appears and by following its subsequent growth with field. In the unperturbed 2DES, the filling factor n=2 (one Landau level filled with both spin-up and spin-down electrons) occurs at 6.1 T. The black spot of Fig. 2 appears at 1.5 T, and grows monotonically as the field is raised. As this growth is observed to occur beyond the field for n=3 in the bulk, we conclude that the spot must first appear when the filling factor at the center of the perturbed region is n=2. The spot emerges at ¼ of the field required for n=2 in the bulk, suggesting that the density at the spot has been reduced to about ¼ of the bulk density.

Fig. 3 displays a series of five images taken at magnetic fields spaced by 0.005 T. The images reveal a startling amount of detail (see caption). One may quickly ascertain that the features all indeed arise from charging structure within the 2D layer resulting from the quantum Hall effect. There is no scenario for electronic structure in the donor layer or on the sample surface having such profound field sensitivity. Moreover, the structures are only seen when the magnetic field is near Hall plateaus. Lastly, features redevelop when the magnetic field is increased by precisely a factor of two, corresponding to the change from $\nu=4$ to $\nu=2$; the path followed by the filaments at 3.000 T (see schematic in Fig. 4) is once again followed by filaments only at precisely 6.000 T (near $\nu=2$). No known physical process would give rise to such periodicity other than the quantum Hall effect.

What is the origin of the contrast in the 2DES? The observed structure may arise due to spatial variation in compressibility of the 2DES, i.e., some regions of the 2DES can absorb more charge then others, or alternatively, due to differences in local charging rates. It is possible to distinguish between these two mechanisms by comparing the contrast in the $Q_{in}$ and $Q_{out}$ images.

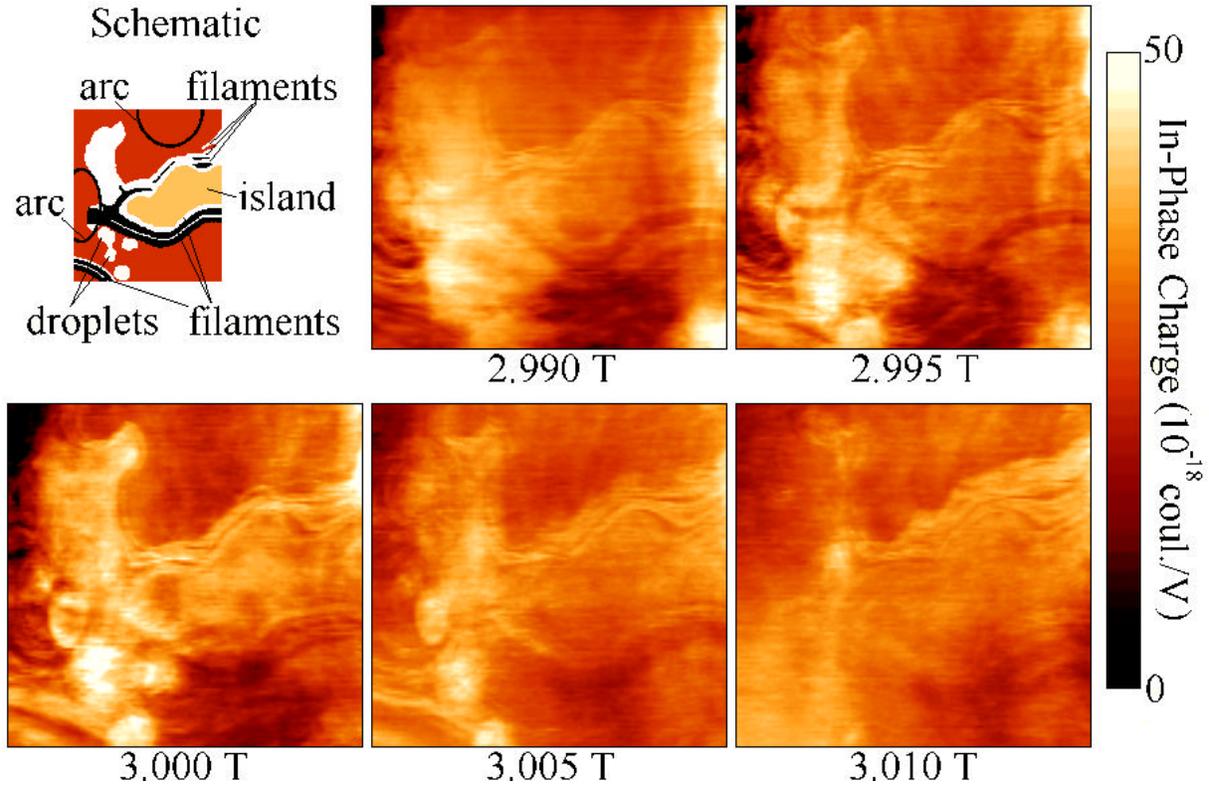

**Figure 3** Sequence of five in-phase SCA images (3.3mm x 3.3mm) for an unperturbed region of the 2DES at n≈4.05 for magnetic fields differing by 0.005 T. Each image was obtained using five hours of signal averaging. Kelvin probe measurements were used to find a region unperturbed by static charge fluctuations. The contact potential was measured with the microscope in Kelvin probe mode, and the tip bias (+0.80 V) was then set to eliminate dc fields between the tip and the 2DES. To enhance our ability to display small scale features, a background plane has been subtracted from each image. This does not alter the images significantly nor create structural artifacts in the images. Finally, despite the great sensitivity of the structures to changes in magnetic field, the structures do not change at all with time. The expected spatial resolution of the probe is 900 Å, consistent with the smallest features displayed in the images. "Filaments" (see schematic) of high or low charge accumulation appear in each of the images. In the 3.000 T image, these filaments are separated by about 1500 Å and meander in unison for a distance of at least two microns in the image. The filaments exist at the boundary between broader regions of high and low charge accumulation. Some filaments appear as broad dark bands that wind all the way across the image. Also, a grouping of high charge accumulating bulbous "droplets" are clearly demarcated in the images at 3.000 T and 3.005 T. The droplets shrink rapidly in size as the field strength is increased. Finally, "arcs" or "loops" are apparent in the images. The upper arc shown in the schematic exists in the image at 2.995 T, and it appears to follow the circumference of a loop of diameter ~1.5 mm.

One can think crudely of the 2DES as an *RC* system (where *R* is the resistance of the layer, and *C* is its self-capacitance). In the experiment, the resistivity of the sample varies as a function of magnetic field, and becomes extremely large at integer filling. At these fields, the roll-off frequency $f_0 = 1/(2\pi RC)$ may approach the excitation frequency (100 kHz). As a result, in narrow intervals of field around integer fillings we detect both $Q_{in}$ and $Q_{out}$ components, i.e., the signal has a nonzero phase shift. However, at fields somewhat higher or lower than this interval no detailed contrast is observed in $Q_{out}$ but much is seen in $Q_{in}$. Figure 4 displays this behavior by showing $Q_{in}$ and $Q_{out}$ images for two different magnetic fields, with the higher field (3.005 T) closer to integer Landau level filling factor.

A more elaborate model[26] treats the 2DES as a distributed *RC* system with no unique roll-off frequency. Consider a region of size *L*, ranging from the size of the sample $L_{max}$ (several mm) to the resolution limit given by $L_{min}$ (900 Å). Given a sheet of uniform resistivity in two dimensions, the effective resistance of this region does not depend on its size, but the self capacitance increases linearly with L. The roll-off frequency depends on the size *L* of the region in question: $f_0(L) = 1/(2\pi RL)$. Thus, $f_0(L)$ increases as *L* diminishes. Therefore, when the roll-off frequency becomes of the order of the measurement frequency, the charging of smaller length scales still behaves as in the low frequency limit, and features charge in-phase with their surroundings. Thus *only* compressibility features can be detected at small length scales.[26]

Why is the contrast due to compressibility variation seen in such a narrow interval of magnetic fields? The simplest answer is that the features may exist only near integer fillings. Another possibility is that the contrast is enhanced near integer fillings because of the lack of screening by the 2DES. This enhancement will occur when the screening radius in the 2DES becomes of the order of the distance to the tip, i.e., in a narrow interval around an integer filling.

We now consider what properties of the QHE could give rise to the observed structures. From measurements of the sample capacitance vs. magnetic field, we know that ν=4 occurs at 3.040 T in the bulk. The droplets (see schematic in Fig. 3) seen at 3.000 T and 3.005 T can be explained as regions of higher electron density than their surroundings, which contain electrons in Landau levels

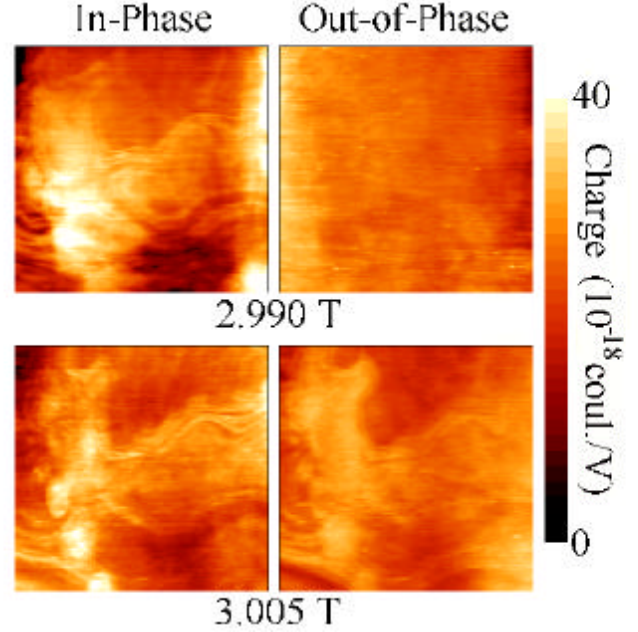

**Figure 4** $Q_{in}$ and $Q_{out}$ images at 2.990 T (top row) and 3.005 T (bottom row). Notice that $Q_{out}$ is nearly featureless at 2.990 T while $Q_{in}$ displays detailed structure. Increasing the field by only 150 gauss to 3.005 T dramatically alters the phase of the image. The overall resistance of the sample increases as the field is raised from 2.990 T to 3.005 T, and the model for charging of the 2DES (see text) explains the observed phase shift. Notice that at 3.005 T, the contrast seen in $Q_{out}$ always has the same sign as that in $Q_{in}$. Based on a simple *RC* charging model, it is possible to show that this observation is consistent with our conclusion that the structures in the images arise from variations in local compressibility and *not* from variation in local charging rates.

ν>4. As the magnetic field strength is increased, the higher Landau levels depopulate, and the droplets shrink in size[27] and eventually disappear altogether.

At lower fields, the droplets are larger and at 2.995 T appear to overlap with a relatively bright "island." We thus surmise that the island is also a region of higher electron density where ν>4. Filaments border this area and thus occur at positions where there is a density gradient. We note that we have never observed filaments which terminate. Either they form closed loops or meander all the way across the images.

To understand the origin of filaments one has to explain clumping of the electron density in a 2DES where the main interaction between electrons is Coulomb repulsion. It is known that in some cases the exchange interaction between electrons can overcome the direct density-density interaction and change the effective interaction from repulsion to attraction. There are several models that use exchange and predict filamentary structures in the quantum Hall system. In particular, Chamon and Wen[28] and MacDonald[29] have described charge structures forming near the edges of a 2DES in the presence of a large density gradient. Recently, Koulakov, Fogler, and Shklovskii (KFS)[30] considered the 2DES containing slightly greater than integer number of filled Landau levels (e.g. the same situation as Fig. 3). Using the Hartree-Fock method, KFS found that the electrons in the higher level tended to form a charge density waves, i.e., a family of equally spaced filaments.

Observing carefully the "island" (see Fig. 3) as a function of field, it is clear that unlike the "droplets," this region does not shrink but rather expands with increasing magnetic field. This at first seems inconsistent with the idea that the island contains electrons in a higher Landau level which depopulates as the field is raised. However, as the field increases, more dark filaments appear toward the center of the island. In general, as the field is increased, the filaments widen into dark and bright bands moving from the center of the island outwards. The images suggest that some regions of higher $\nu$ depopulate in a different fashion than considered previously. In the island, the density of electrons in the higher Landau level is reduced by "stretching" the area. The electron density in the island can be thought of as an accordion, and the spacing between bands of high electron density increases with increasing magnetic field.

We have shown that the SCA technique has a profound capability to look into systems that have been inaccessible to imaging with comparable resolution. Even though it is not presently possible to determine the exact origin of the structures, there is no question that they actually *exist*. The observed shapes, rapidly evolving in magnetic field, directly mirror the electronic structures in the highly correlated Hall fluid. The range of experimental conditions which produce such shapes, the importance of disorder, and the nature of the fractional Hall regime will be objects of further study.

## Acknowledgements

We gratefully acknowledge technical help from A. Cohen, E. Atmaca, M. Brodsky, H.B. Chan, A. Folch, S. Heemeyer, A. Shytov, D. Silevitch, and N.B. Zhitenev and thank B.I. Halperin, H.F. Hess, R.B. Laughlin, P.A. Lee, and B.I. Shklovskii for helpful discussions. This work was supported by the Office of Naval Research, the Packard Foundation, JSEP, and the National Science Foundation DMR.